\documentstyle[proceedings,numreferences,epsf]{crckapb}

\input epsf

\begin{opening}

       \title{TOPICS IN QUANTUM COMPUTERS}

       \author{D.~P.~DiVINCENZO}
       \institute{IBM Research Division\\
		  T. J. Watson Research Center\\
                  Yorktown Heights, NY 10598 USA}
\end{opening}

\runningtitle{Quantum Computers}
\makeindex

\begin{document}
       \begin{abstract}
I provide an introduction to quantum computers, describing how they
might be realized using language accessible to a solid state physicist.  A
listing of the minimal requirements for creating a quantum computer is
given.  I also discuss several recent developments 
in the area of quantum error 
correction, a subject of importance not only to quantum computation,
but also to some aspects of the foundations of quantum theory.
       \end{abstract}

\section{What is a quantum computer?}

I don't think that I will spend many words here saying why there has
been a considerable growth of interest in the last couple of years in
the subject of quantum computation.  There has been a spate of
reviews\cite{PT,Div2,EJ}, semi-popular articles\cite{SciAm}, and press
accounts\cite{Gla} giving, on the whole, a very good overview of the
subject.  At some level, the recent interest simply arises from the
very traditional movement of computation into ever more miniature
worlds, and what could be more miniature than the world of the single
quantum?  At another level, though, interest has arisen because the
rules of quantum dynamics changes the rules of computation
itself\cite{G9}, in ways which we are still working to understand.
You probably can't factor large numbers with any computer following,
at the level of the logical operations, the laws of classical
mechanics (which is to say, every computer ever operated up until
now); with a computer, or a computation, obeying the laws of quantum
dynamics, you just might be able to factor\cite{Shor,EJ}.  This has
drawn the attention of both the practical, problem-solving world, as
well as that of those interested in further exploring and
understanding the foundations of quantum theory itself.

So, since we are all solid state physicists here at Curacao, and all
know a thing or two about quantum physics, let me immediately give
a fairly sophisticated run-down of the minimal requirements for any
quantum system to be a quantum computer.  As you will see, the entry
fee is pretty steep, which provides at least one good reason why
all sorts of people aren't already putting together their quantum
processors.  Anyway, here goes:

\subsection{Five Requirements for Quantum Computing}
 
1) The degrees of freedom required to hold data and perform computation
should be available as dimensions of the Hilbert space of a quantum
system.  This quantum system should be more-or-less isolated from 
its environment.  (More about this shortly.)  Also, this Hilbert 
space should be precisely enumerable, for example, I should be able to
say, ``the quantum system consists of 49 spin-3/2 states on the molecule,"
or something of this sort.  It won't do to be able to make only statistical
statements about the number of degrees of freedom, as is often the case
in solid state physics (e.g., ``the quantum dot contains $100\pm 5$ 
electrons").  No, the Hilbert space must be {\em precisely} delineated.

Now, it is furthermore very desirable for the Hilbert space to be
decomposable into a {\em direct product form}.  This rather
formidable-sounding requirement is actually quite natural for a
multi-particle quantum system.  In the example mentioned a moment ago,
the Hilbert space decomposes into a product of 49 parts, each of
dimension $2(3/2+1)=5$.  Note that this means that the dimension of
the Hilbert space of the entire quantum system is $5^{49}$, a rather
large number.  When the Hilbert space of the individual particle is
two-dimensional (e.g., for a spin-1/2 particle), we term this object a
quantum bit or {\em qubit}, as its role in quantum computation is
analogous to that of an individual bit in an ordinary computer.  Now,
it is not absolutely necessary that the Hilbert space have this
direct-product form; the Hilbert spaces of collections of
indistinguishable particles do not have this form because of
(anti-)symmetry constraints.  But it is essential that the size of the
Hilbert space grow exponentially fast with the size of the system.
For this, a multi-particle system is necessary; the Hilbert-space
dimension of a one-particle system grows only algebraically with
system size.

2) Another requirement of quantum computation is that it must be possible
to place the quantum system in a fiducial starting quantum state.  This
state can be very simple, as in ``all spins down" for a collection of
spins; thus, this requirement will be satisfied if it is possible to
{\em cool} the quantum system to its ground state.  This may be trivial
--- if the qubits are embodied by a ground and excited state of an atom,
room temperature may be quite cool enough; this cooling requirement may
also be very high tech --- cooling atoms in a trap to their motional
ground state, or cooling nuclear spins to their ground states, may
require nano-Kelvins.  Of course, the experimentalists have not been
remiss in achieving these sorts of conditions lately.  I do not think
that this ``initial state preparation" requirement will be the most
difficult one to achieve for quantum computation.

3) Here is a pretty tough requirement.  The quantum system to be used
as a quantum computer must be to a high degree {\em isolated} from
coupling to its environment.  This isolation requirement is linked up
with the precision required in quantum computation: if the state of
the computer at some instant is ideally supposed to be the state
$\Psi$, then the actual state after one clock cycle, $\rho$ (a {\em
density matrix}), should differ from $\Psi$ by only a small amount:
\begin{equation}
\langle\Psi|\rho|\Psi\rangle\ge 1-\epsilon.\label{eq1}
\end{equation}
One way (not the only way) that the state of the system can depart
from $\Psi$ is by evolving into a state in which the joint quantum
state of the computer and that of its environment become correlated
through interaction, a state of affairs described as {\em
entanglement} by Schroedinger in 1935\cite{Schr,Schr2}.  When the joint
state is entangled, the state of the system alone must be described as
a mixed state or a density matrix; in mesoscopic physics, we would say
that the quantum system has begun to travel down the road to
decoherence or phase-breaking.

By the way, the issue of how big an $\epsilon$ is tolerable in quantum
computation is probably {\em the} active question among theorists in
this field today, as it is all tied up in the question of error
correction and fault tolerance in quantum computation.  There is good
news and bad news on this front: error correction {\em is} possible in
quantum computation, so finite $\epsilon$ is perfectly tolerable.  The
bad news is, at least at the moment, it is not known whether a very
{\em big} $\epsilon$ is tolerable.  The protocols which are currently
understood\cite{FT} start to get into trouble when $\epsilon=10^{-6}$,
give or take a few orders of magnitude.  But there may be schemes in
which $\epsilon=0.1$ will be perfectly all right; we just don't know,
and I certainly hope that we manage to figure it out.

4) The next requirement is at the heart of quantum computation: It
must be possible to subject the quantum system to a controlled
sequence of unitary transformations.  All of our quantum algorithms
are expressed in terms of such sequences.  It is also required that
these unitary transformations can be made to act upon specified pairs
of qubits, or other small collections of qubits.  This fits
hand-in-glove with the requirement 1) that the Hilbert space be made
of a direct product of the spaces of individual small systems or
particles.  Note that the necessary effect of the two-bit quantum
transformations (which are referred to as {\em quantum logic gates})
is that they produce entanglement {\em between} qubits of the quantum
computer.  Notice that entanglement between different parts of the
quantum computer is good; entanglement between the quantum computer
and its environment is bad, since it corresponds to decoherence.
 
There are various ways in which such controlled unitary transformations
may be achieved.  In most of the ways which people are thinking about now,
it is attained by subjecting the system to a particular time-dependent
Hamiltonian over a fixed length of time; the resulting unitary operator
is a function of the Hamiltonian according to the usual time-ordered
product expression:
\begin{equation}
U=T\exp(i\int H(t)dt)
\end{equation}
(I won't get into the time-ordering mumbo-jumbo here.)  This $H(t)$
may be imposed via a time-varying magnetic field (as in NMR), a strong
(i.e.  classical) time-varying optical field (as in laser
spectroscopy), or by the physical motion of a massive (again,
therefore, classical) object (viz., the tip in an atomic-force
microscope).  Such $H(t)$'s are in fact ubiquitous in experimental
physics, although making these operations selective at the
single-quantum level is not.

Again, there is what appears at the moment to be a pretty stringent
precision requirement on these unitary operations.  In the natural
metric in the space of unitary operations, the distance between the
specified and the actual unitary transformation should be less than
$\epsilon$, and the present constraint on $\epsilon$ is about the
same as in the discussion in item 3); only $\epsilon$'s less than about
$10^{-5}$ can presently be considered ``safe''.

5) Last, but not least, it is necessary that it be possible to subject
the quantum system to a ``strong'' form of measurement.  When I say
``strong'' I simply refer to the kind of quantum measurement that we
learned about in our textbooks: the measurement determines which
orthogonal eigenstate of some particular Hermitian operator the
quantum state belongs to, while at the same time projecting the
wavefunction of the system irreversibly into the corresponding
eigenfunction.  The standard example of such a measurement is the
Stern-Gerlach experiment in which the z-component of a spin-1/2
particle is projected into one of its two eigenvalues.  While this is
straight out of the textbooks, it is unfortunately unlike what a lot
of actual quantum measurements in the laboratory actually consist of,
being more of a ``weak'' variety.  I will not give a complete
discussion of what a weak measurement is, I could point the reader to
the textbook of Peres\cite{Per} for this discussion.  Basically, the
idea is that the individual quantum system, say a single spin-1/2
system, might interact very weakly with the measurement apparatus,
such that the probability that the apparatus registers ``spin up'' is
only very weakly correlated with the actual wavefunction amplitude for
the spin to be up.  To be more quantitative, there exist weak
measurements on a state $\Psi=a|\uparrow\rangle+b|\downarrow\rangle$
which register ``spin up'' with probability
\begin{equation}
p_{up}=\frac{1}{2}+\delta\left (|a|^2-\frac{1}{2}\right )\label{weak}
\end{equation}
for arbitrarily small $\delta$.  These small-$\delta$ measurements
are ``weak'' in the sense that after such a measurement the quantum
state of the system has been disturbed hardly at all; on the other
hand, hardly any information has been gained by the measurement about
the state of the spin.  In many areas of experimental physics a weak
measurement is all that you can do, simply because it is not presently
known how to make the coupling to individual quantum systems sufficiently
strong; in NMR and in most solid-state spin systems, the
experiments are insufficiently sensitive to detect the state of individual
spins, although there is a continuing push in that direction.  I would say
that electrical measurements of mesoscopic quantum structures are also
weak, in the sense that the course of each individual electron passing
through the device is typically not determined.  (This is probably
not the case for certain measurements in single electronics.)  

In many cases weak measurements are very satisfactory for learning
a great deal about the quantum properties of systems, because they
can often be done on macroscopically large ensembles, involving either
many replicas of the same quantum system (very typical in NMR), or
many identical runs of the same quantum measurement (as in normal
electron transport).  By averaging over such ensembles, a very
good knowledge of $a$ in Eq. (\ref{weak}) can be obtained, no matter
how small $\delta$ is.  However, these weak measurements do {\em not}
satisfy the requirements in quantum computation, at least so far as 
we have presently formulated them.

\section{Brief survey of experimental systems}

I believe that with this long-winded set of five desiderata, a
reasonable evaluation of any proposed quantum computer implementation
can be made.  Let me illustrate this with two brief examples, both
of a nano-solid state character, which show that the very act of
performing this evaluation points to a lot of physics which we would
like to know about these systems, about which we are presently pretty
ignorant.

\subsection{\em Atomic Force Microscope.}  

This is a gedanken apparatus, shown in cartoon form in Fig. 1, which I
proposed some time ago\cite{Div3,Div2} for doing quantum computation.
As will be seen, there are in fact severe problems with using it as a
quantum computer; but indeed, that is what we are supposed to use the
magic five criteria to reveal!

\begin{figure}
\vspace{5cm}   
\caption{My cartoon of at atomic force microscope concept for a 
quantum computer, described in the text.  See also
\protect\cite{Div2}.}
\label{fig1}
\end{figure}

1) The idea of precisely controlling the computational Hilbert space
is tied up in the precise atomic design implied by the cartoon for the
tip and the surface of the instrument.  Finding a spin within a
solid is not a problem; indeed there are spins and other quantum
states galore.  The idea is to set up the system so that only the
ones you want are really ``available''.  A qubit may be made unavailable
if the energy to change its state is much larger than the energy actually
available in the experiment; in solid state language, the idea is to
``open a gap'' for all the excitations that you don't want to happen.
In the cartoon, these means that we envision an insulating material
such as intrinsic x-Si at low temperatures, so that there are no
low-lying excitations available.  Furthermore, the nuclei are all
chosen so that no nuclear spin excitations are available (here the
``unavailability''
is particularly clear, since the energy to create a nuclear 
excitation would be
somewhere in the MeV range).  So there are no low-lying states, 
{\em except} for the
proton spin of the H atoms terminating the tip and various sites on
the surface.  Thus, this is envisioned as a proton-NMR quantum computer.
It places some requirements on atomic scale atomic engineering (the
avoidance of any defects within or on the surface of the crystal, the
avoidance of surface states, precise isotopic control) which are at
the moment pie-in-the-sky, although conceivable as an extension of the
considerable actual progress with atomic placement and manipulation in
recent years.

2) The preparation of such a nuclear spin system into a pure state
would require some cooling techniques considerably beyond the current
state of the art.  Since the energy scales for nuclear spins are so
small, the temperature required to have a substantial ground-state
population is below $10^{-3}$K or so, for spins in a 1T field.  Of
course such temperatures can be attained in many types of experiments,
but I know of no effort to achieve them in an AFM.  There are
techniques which are specifically adapted to bringing nuclear spins to
low temperatures\cite{Chupp}, involving coupling by spin resonance
techniques to other spins (typically electron spins).  These
approaches could conceivably be applied in an AFM, although no one has
undertaken to do so up until now.

3) There are also many unknowns about the question of how well isolated
such proton spins could be from the environment.  The small energy scale
of the nuclear spin is an advantage, as it reduces the phase space available
for the emission of excitations (phonons, say) into the crystal.  Other
stray spins, from defects or thermally activated electrons, say, would
be a concern.  I would not want to try to predict the decoherence time
attainable in such an instrument from a purely theoretical approach;
ultimately, it is, I think, an experimental question.  There are
localized spins in solids which are known to have lifetimes in excess
of 1000 sec\cite{Castle}.

4) We know that, almost by definition, an AFM is capable of placing
the tip with respect to the surface with something approximating 
atomic-scale accuracy.  This presumably means that one can turn on
and off the spin-spin interaction Hamiltonian between a spin at the
tip of the AFM and a selected spin at the surface, at least with
O(1) accuracy.  We normally do not think that O(1) accuracy is anywhere
near good enough, as discussed above we are more comfortable with O($10^{-6}$)
accuracy.  I think that it is completely unknown whether the AFM's
positioning accuracy could be made that good, or whether protocols
for quantum computation can be devised in which a low level of
positioning accuracy is acceptable.  Not that even the possibly
crude atomic selectivity of the AFM is not available at all in many
other experimental techniques in solid state physics.

One reason why O(1) accuracy of positioning may be enough is that the
principal purpose of this interaction as I have envisioned it is to
shift the resonance frequency of the selected pair of spins so that
their state can be manipulated independently from the other spins in the
system.  One can tailor pulse sequences in NMR which will accurately
perform some desired unitary transformation on the pair of spins in
such a way that the result is insensitive to the exact resonance
frequency, so long as it falls in a given range.

Another problem which spans both the issues of isolation and
controllability, which I am grateful to Prof. Jon Machta for bringing
to my attention, is this: the surface will exert a force on the tip,
which will be dependent upon the state of the two spins.  In principle
this means that the trajectory of the tip $\vec r(t)$ is dependent
upon the state of the spins.  This can be very bad for quantum
computation, since it spoils the desired isolation of the quantum
state from the environment; if the quantum state of the spins becomes
entangled with the state of motion of the tip, the quantum computation
state will be effectively decohered.  Fortunately the situation is not
as bad as it looks, for several reasons: first, it is reasonable to
take the tip not to be in an eigenstate of position $\vec r$, but in a
coherent state involving some reasonable degree of uncertainty in the
position $\Delta \vec r$ and momentum $\Delta \vec p$
of the tip.  Since the tip is rather massive and
the nuclear spin forces are rather small, the deflection of the tip
due to these forces will in fact quite small, and it would be
reasonable to imagine that this deflection is well within the
position uncertainly $\Delta \vec r$ 
of the tip.  In this case, the entanglement of
the motion of the tip with the quantum computation is quite small.  In
fact this is the same form of argument that is needed to say why the
radio-frequency spectroscopy, which involves absorption and
emission of phonons, does not represent a loss of coherence to the
environment.  The point is that the radio-frequency electric and
magnetic fields are coherent states involving considerable
uncertainty in the photon number, so that one photon more or less does not
change the quantum state of the r.f. field appreciably.

There is one other subtlety concerning the spin forces on the tip.
As noted, the trajectory is changed in no significant way by
this interaction; however, the interaction does apply an impulse
to the tip, dependent on the state of the spins $s_1$ and $s_2$,
\begin{equation}
\delta \vec p(s_1,s_2)=\int \vec F(s_1,s_2,t) dt\label{smallf}
\end{equation} 
which results in a subsequent change of the phase
evolution after this impulse:
\begin{equation}
\exp(i \delta \vec p(s_1,s_2)\cdot\vec r(t)/\hbar)
\end{equation}
($\vec r(t)$ is the subsequent trajectory of the tip).  Note that so
long as knowledge of this phase evolution is retained (requiring, for
example, quite accurate knowledge of the whole trajectory 
$\vec r(t)$ of the tip),
this effect is not decohering; it merely appends a definite phase to
each part of the quantum computer state, which must be accounted for
in the bookkeeping of doing the prescribed unitary transformations 
properly.

5) Finally, the situation for actually performing a strong 
measurement of the
quantum state of individual spins in the AFM is rather hopeful, in the
sense that this has already been recognized as a valuable thing to do,
and several researchers have been carrying out a whole experimental
program to try to do such measurements\cite{Rugar}.  They are very
hard, indeed, and have not yet been accomplished --- they involve, in
fact, the self-same minute forces mentioned in Eq, (\ref{smallf}),
which, I have already noted, are very hard to entangle with an
external variable like the tip position.  However, if you really {\em
want} this entanglement to occur, a few different tricks are available
to you to amplify the tiny entangling tendency: the tricks involve
keeping the tip in contact with the spin to be measured for a long
time, and arranging that the force oscillate in time (flipping the
spin back and forth with appropriately timed tipping pulses) so as to
excite a mechanical resonance of the tip.  Estimates indicate that
under favorable circumstances this strategy can achieve single-spin
sensitivity, although we will see if the experiments ever actually
manage it.

\subsection{\em Josephson junction device}

Let me consider, in even less detail than above, the five criteria for
quantum computation as applied to Josephson junction systems.  I will
not pretend that I seriously know how to perform this evaluation
realistically, but I think that other participants at this school
might, prompted by my feeble attempt, be able to do a much better
job of it.

1) The Hilbert space we would have in mind would describe the
quantized states of the superconducting phase.  The fact that the
phase is a ``macroscopic quantized variable'' is by no means an
unalloyed advantage (see below), but at least it offers the
possibility of sculpting a Hilbert space by a suitable design of an
electric circuit.  What would be the good way of doing this sculpting
I do not really know; one possible approach, inspired in my mind by
some comments by Prof. Mooij, is to use the quantized states of
position of superconducting vortices\cite{Mooij} as the relevant
degree of freedom.  So, the two basis states of a qubit may be a
vortex positioned in one superconducting loop or another neighboring
one (with a Josephson junction in between).  It would be necessary to
have ``macroscopic quantum coherence'' in that this ``macroscopic''
vortex would have to be capable of existing in a quantum superposition
of positions.

2) Concerning state preparation, it seems possible that, by suitable
external application of supercurrents, the potential profile of the
superconducting state may be biased such that, for example, the
lowest energy state is the vortex in one particular position.
Presumably this would result in the preparation of a pure state,
although one has to worry about thermal excitations of the vortex
into other states within the same ring, say.  There are a lot of
unknowns on this score.
 
3) Again, on the issue of the isolation of the system from its
environment, we would quickly enter the realm of guesswork.  The
experiments to pin this down would correspond rather closely
to those which have been attempted for years to
document the occurrence of MQT and MQC (macroscopic quantum
tunneling and coherence) in these systems\cite{Rouse}.  Such experiments
would have to work before we could say that the coupling to
the environment is sufficiently under control to embark on quantum
computation experiments.

4) The manipulation of the time evolution of the system,
by applying desired time dependent Hamiltonians to the system, is
probably the most achievable of the five criteria.  The effective
Hamiltonian of the quantized superconducting phase (or of the
conjugate number operator) contains a variety of parameters which
are determined by the macroscopic condition of the circuit, for
instance the capacitance or inductance of various circuit elements,
or the value of various externally applied supercurrents in the circuit.
All of these, in principle, could be employed to implement quantum gates.
It would make sense to explore protocols for doing so if the other
criteria for quantum computation were closer to being sorted out.

5) Again, the fact that the system is more-or-less macroscopic
makes it easy to envision various kinds of measurements, of local
magnetic flux, of voltage, etc., being performed reliably on the
system.  The headache comes from the fact that when the means for
performing these measurements are put into place, one must ask 
whether they form a potentially destructive part of the environment.
The mere act of refraining from viewing the display of a voltmeter
does not change the fact that the measurement setup is collapsing the
wavefunction of the computation in an undesirable way.  The point is
that there must be some way of coupling and decoupling the measurement
with the quantum system at any desired moment, as is routinely done
by applying appropriate resonant r.f. fields in NMR.  I do now know
how this criterion is to be satisfied, although I believe there is
some consideration of this issue in the work of Leggett\cite{Leg}.

---

Any of you who work in superconductivity who have just read this must
be convinced that quantum computation is just some wild-eyed notion
that bears no relationship to reality at all.  But the remarkable (or
perhaps depressing) fact of the matter is that in other fields of
experimental physics workers have a certain degree of confidence that
the five criteria which I have laid out are within the grasp of their
experimental technique.  The most notable example is the consideration
of the {\em linear ion trap} which has been put forward in the work of
Cirac and Zoller\cite{Cirac}.  I can succinctly summarize their
proposals according to my five-point plan, thus: 1) Hilbert space:
very precisely understood --- it is spanned by the energy levels of
the isolated ions, combined with the phonon modes of the ions in the
trap.  2) Cooling: done --- the capability of laser cooling into the
ground state was demonstrated a couple of years ago (for single ions,
at least; cooling in multi-ion traps still has a ways to go).  3)
Isolation: pretty good --- the coherence times of the ion levels is
unmeasurably long, the phonon lifetimes are adequately long to make a
start of quantum computing (although these lifetimes are a few orders
of magnitude shorter than are presently understood theoretically).  4)
Unitary operations: done, with very high precision, by laser
spectroscopy.  (There is a lot of detailed atomic physics that is
being argued about now on how precisely these state transformations
can actually be effected, but the arguments are at the
1-part-in-$10^4$ level.  Also, the relevant precision spectroscopy has
not been demonstrated for the multi-ion experiments.)  5) Measurement:
Perfect --- The laser-induced fluorescence technique has 100\% quantum
efficiency, and can be turned on and off at will.

Despair may be the reaction of those of you reading this who had hopes
that a solid-state implementation of quantum computing might be
possible, or at least competitive with what can be done in atomic
physics.  But please don't let the apparent tone of finality in the
survey of prospects for the ion trap, nor the tone of uncertainty and
pessimism in the survey of the superconducting implementation, deter
you.  For one thing, the ion trap has more problems than my survey
revealed.  First, the ion-trap scheme is not very extendible; the most
ions which have ever been trapped in such an apparatus is in the
neighborhood of 40; moreover, all the various cooling and
spectroscopic tricks have been performed presently only on {\em
single} trapped ions\cite{Monroe}.  The potential extendibility of a
solid-state system, if can be made to work at all, would seem to be
much greater.  Second, and most obvious, I may have stupidly missed
some brilliant way of implementing quantum computation in
superconducting circuits which does not suffer from any of the
problems which my thoughts entail.  Let not my lack of brilliance
stand in the way of yours.

\section{Current concepts in error correction}

Having provided a grand tour in the last section of the requirements
for quantum computation, I propose to get down in the trenches a
little bit, and discuss how we are trying to solve some particular
problems on the subject of protecting quantum computation from
errors\cite{FT,Shor9,us}.  This is primarily a subtopic of criterion
3) of my list above, since it has to do with how well the quantum
computer can survive interactions with its environment.  It also bears
upon criterion 4), in so far as it also provides a prescription for
how to tolerate (slightly) inaccurate quantum-gate operations (that
is, the unitary transformations involved in quantum computation).

The following discussion will have no pretension to be a complete
description of what is going on now on the subject of error
correction.  It is intended as an introduction of what I consider to
be an interesting subset of the current work on the subject.

\subsection{Description of errors}

The description of the interaction of a quantum system with its
environment has been a subject for theorizing for many decades now,
and can be as complex as you please.  Let me bring up a number
of elementary points about this description, which will be the
only points which are needed for quantum error correction.

Suppose that we begin with a single two-level quantum system, and that
it starts out at time $t=0$ in a pure initial state $\Psi$.  If this
system then begins to interact with its environment, the state of the
system plus its environment undergoes some joint evolution.  Viewing
the evolution just from the point of view of the two-state system, the
initial state $\Psi$ evolves into a {\em mixed state}, which may
either be thought of as an incoherent statistical mixture of an
ensemble of new pure states, or it may simply described by a {\em
density matrix} $\rho$\cite{Per}.  The linear operator which specifies
the time evolution of such an open system is termed a {\em
superoperator}\cite{KL4,Shoe}; 
this jargon, I believe, is meant to distinguish it
from an ordinary (unitary) time-evolution operator describing the time
evolution of a closed quantum system.

A very useful mathematical description of the superoperators is in
terms of the so-called {\em operator sum representation}, in which
this time-evolution is described as follows:
\begin{equation}
\rho(t)=\sum_{i=1}^4A_i\rho(0)A_i^\dagger,\label{osr}
\end{equation}
\begin{equation}
\sum_{i=1}^4A_i^\dagger A_i=1,\label{const}
\end{equation}
\begin{equation}
\rho(0)=|\Psi\rangle\langle\Psi|.\label{init}
\end{equation}
The action of {\em every possible} environment is completely describable
by the four operators (2$\times$2 matrices) $A_i$; 
these matrices may take any form consistent
with the completeness condition Eq. (\ref{const}).  The $A$ matrices
also come up in the ``ensemble'' description of the effect of the quantum
environment; we would say that the state $\Psi$ evolves into the
ensemble of four states
\begin{equation}
|\Psi\rangle\rightarrow\{A_1|\Psi\rangle,A_2|\Psi\rangle,A_3|\Psi\rangle,
A_4|\Psi\rangle\}\label{ensem}
\end{equation}

\subsection{Error Correction Conditions}

Now the ideal of quantum error correction is that it should be possible
to subject the corrupted quantum
state to some process 
${\cal R}$ which combines measurement and unitary transformation
that brings any corrupted state back to the state without errors:
\begin{equation}
A_i|\Psi\rangle\stackrel{\cal R}{\rightarrow}|\Psi\rangle.\label{desire}
\end{equation}
Obviously it is impossible to satisfy Eq. (\ref{desire}) for any
arbitrary $\Psi$ and error process $A_i$.  The remarkable fact discovered
recently is that, if the subspace in which the state $\Psi$ lies in is
restricted, and the set of allowed error operators $A_i$ is also restricted
(in a physically reasonable way, as it turns out), then the desire expressed
by Eq. (\ref{desire}) {\em can} be satisfied.  I will review here the
derivation of the conditions to be satisfied in order that this
{\em quantum error correction} process be possible, which may be found
in our paper\cite{us} and in a number of others\cite{KL4,Niels,EkMac}.

The basic idea is just to formalize Eq. (\ref{desire}) a little more 
carefully.  Let me specialize to the case where the error-protected states
$\Psi$ are to lie in a two-dimensional subspace of the Hilbert space of
$n$ spins --- the jargon for this is that we will use $n$ qubits to
store one qubit robustly.  Introducing an orthogonal 
basis for this two-dimensional
subspace in an obvious way, we will write the general $\Psi$ as
\begin{equation}
|\Psi\rangle=a|v_0\rangle+b|v_1\rangle
\end{equation}
with arbitrary complex coefficients $a$ and $b$.  Now, the error
correction process ${\cal R}$, involving some measurements and unitary
operations, can always be thought of as an entirely unitary process
${\cal R_U}$ involving a larger system, the Hilbert space of $\Psi$
along with what we will term an {\em ancilla} Hilbert space.  We will
imagine that the ancilla, which must be under the control of the
experiment in the sense of criterion 1) of the first section, is
always preset to a standard state $|0\rangle$.  Now, we may restate
the requirement, by saying that there should exist a unitary
transformation ${\cal R_U}$ which performs the mapping
\begin{equation}
||A_i|\Psi\rangle|0\rangle||
\stackrel{\cal R_U}{\rightarrow}|\Psi\rangle|a_i\rangle.\label{bigcon}
\end{equation}
(Here the $||...||$ indicates that the state may have to be
normalized first.)
The operation ${\cal R_U}$ is to restore the state as indicated
for all error processes $A_i$ which are to be corrected.
As indicated here, the ancilla will end up in some new state $a_i$ which
is dependent on $A_i$; the main content of this equation is that whatever
the final ancilla state, it must be in a product state with the system,
and the system's state should be restored to its original form, $\Psi$.

Now, by linearity, Eq. (\ref{bigcon}) should apply to each basis state
separately:
\begin{equation}
N_0^iA_i|v_0 \rangle|0\rangle\stackrel{\cal R_U}{\rightarrow}|v_0\rangle|
a_i\rangle\label{map}
\end{equation}
\begin{equation}
N_1^iA_i|v_1 \rangle|0\rangle\stackrel{\cal R_U}{\rightarrow}|v_1\rangle|
a_i\rangle\label{map2}
\end{equation}
I have now introduced explicitly the normalization constants
$N_{0,1}^i$.  Now, it is relatively easy to show that this
normalization factor must be independent of $\Psi$ in order that error
correction work: $N_0^i=N_1^i=N^i$.  I will leave this as an exercise
for the reader to work out (or look up\cite{us}), in order to move on
to the more physically illuminating part of the proof.

The real point is this: in order for the map required in
Eqs. (\ref{map}) and (\ref{map2}) to be unitary, it must preserve the
inner product between any pair of states.  Therefore, the derivation
just consists of writing down all the distinct before-and-after inner
product conditions.  First, the inner product of two states from
Eq. (\ref{map}) with different error operators $i$ and $j$ gives:
\begin{equation}
\langle v_0|A_i^\dagger A_j|v_0\rangle=\frac{1}{(N^i)^2}\langle a_i|a_j
\rangle.\label{v0}
\end{equation}
Doing the same for $v_1$, Eq. (\ref{map2}), gives
\begin{equation}
\langle v_1|A_i^\dagger A_j|v_1\rangle=\frac{1}{(N^i)^2}\langle a_i|a_j
\rangle.\label{v1}
\end{equation}
The right-hand sides of Eqs. (\ref{v0}) and (\ref{v1}) are equal, giving
the first conditions for error correction:
\begin{equation}
\langle v_0|A_i^\dagger A_j|v_0\rangle=
\langle v_1|A_i^\dagger A_j|v_1\rangle.\label{cond1}
\end{equation}
The other error correction condition is given by taking the inner
product of two vectors from Eqs. (\ref{map}) and (\ref{map2}); since
$|v_0\rangle$ and $|v_1\rangle$ are orthogonal:
\begin{equation}
\langle v_1|A_i^\dagger A_j|v_0\rangle=0.\label{cond2}
\end{equation}

\subsection{Use of Canonical Errors for General Case}

So, Eqs. (\ref{cond1}) and (\ref{cond2}) specify the quantum states
$|v_0\rangle$ and $|v_1\rangle$ which can be error-corrected after
being subject to one from among the set of errors $\{A_i\}$.  This may
seem to be not generally very useful because each quantum mechanical
system will have some environment, described by some very particular set
of $A_i$'s.  Nevertheless, there are some generic error-protected
spaces $|v_{0,1}\rangle$ which will be correctable against a whole class
of errors.  To explain this I have to introduce the idea of a 
{\em canonical error set}.  Consider first the environment of
a single qubit.  Note that a complete basis for $2\times 2$ matrices
is given by the set of matrices
\begin{equation}
\begin{array}{ll}E_0=\left(\begin{array}{rr}1&0\\0&1\end{array}\right)&
E_1=\left(\begin{array}{rr}1&0\\0&-1\end{array}\right)\\ \ &\ \\
E_2=\left(\begin{array}{rr}0&1\\1&0\end{array}\right)&
E_3=\left(\begin{array}{rr}0&1\\-1&0\end{array}\right).\label{canon}
\end{array}
\end{equation}
Suppose that we seek a ``generic'' correctable subspace of $n$ qubits
for which we require that error processes which involve one of the
``canonical'' errors of Eq. (\ref{canon}) acting on {\em only one}
out of the $n$ qubits.  To take a specific example, if $n=5$ the
error operators for which the error-correction conditions Eq. (\ref{cond1})
and (\ref{cond2}) are to be satisfied should be these sixteen:
\begin{equation}
{\small
\begin{array}{lll}
A_i^{canon.}\propto \{& & \\
E_0^{(1)}E_0^{(2)}E_0^{(3)}E_0^{(4)}E_0^{(5)},& & \\
E_1^{(1)}E_0^{(2)}E_0^{(3)}E_0^{(4)}E_0^{(5)},&
E_2^{(1)}E_0^{(2)}E_0^{(3)}E_0^{(4)}E_0^{(5)},&
E_3^{(1)}E_0^{(2)}E_0^{(3)}E_0^{(4)}E_0^{(5)},\\
E_0^{(1)}E_1^{(2)}E_0^{(3)}E_0^{(4)}E_0^{(5)},&
E_0^{(1)}E_2^{(2)}E_0^{(3)}E_0^{(4)}E_0^{(5)},&
E_0^{(1)}E_3^{(2)}E_0^{(3)}E_0^{(4)}E_0^{(5)},\\
E_0^{(1)}E_0^{(2)}E_1^{(3)}E_0^{(4)}E_0^{(5)},&
E_0^{(1)}E_0^{(2)}E_2^{(3)}E_0^{(4)}E_0^{(5)},&
E_0^{(1)}E_0^{(2)}E_3^{(3)}E_0^{(4)}E_0^{(5)},\\
E_0^{(1)}E_0^{(2)}E_0^{(3)}E_1^{(4)}E_0^{(5)},&
E_0^{(1)}E_0^{(2)}E_0^{(3)}E_2^{(4)}E_0^{(5)},&
E_0^{(1)}E_0^{(2)}E_0^{(3)}E_3^{(4)}E_0^{(5)},\\
E_0^{(1)}E_0^{(2)}E_0^{(3)}E_0^{(4)}E_1^{(5)},&
E_0^{(1)}E_0^{(2)}E_0^{(3)}E_0^{(4)}E_2^{(5)},&
E_0^{(1)}E_0^{(2)}E_0^{(3)}E_0^{(4)}E_3^{(5)}\}
\end{array}}\label{bigarray}
\end{equation}
Here $E_j^{(i)}$ refers to error $j$ on the $i^{th}$ qubit.  Of course,
to satisfy the completeness condition Eq. (\ref{const}) these sixteen
operators would have to be multiplied by some constants; but these
constants have no bearing on the satisfaction of the error correction
conditions (\ref{cond1}) and (\ref{cond2}), so we can safely ignore 
them.

Now of course the reason I use the example $n=5$ is that in fact we,
and others, have found vectors $v_0$ and $v_1$ that
are perfectly correctable when subject to this
restricted error set.  
Below I will present a complete description of this five-bit
code after I have introduced a good general strategy for finding such
codes.
There is already coming to be a vast literature
on these and related ``codes'', group theoretic strategies for
constructing them, related theorems on the capacity of noisy quantum
channels, quantum gate implementations of the error correction
process, applications of these codes for making quantum computation
fault tolerant; all
of these results are vastly important, but I will leave most of them
for the reader to look up elsewhere.  
For I want to come back to the issue with
which I started, namely, what do the existence of these quantum
error-correction codes for these kinds of ``canonical'' error sets as
in Eq. (\ref{bigarray}) have to say about the case of the ``generic''
environment with arbitrary error operators $A_i$?

The most general answer to this question is, ``nothing.''  The most 
generic environment, which introduces correlated errors among the 
various qubits, bears no resemblance whatsoever to the canonical error
set in Eq. (\ref{bigarray}).  However, if we consider errors which are
non-generic in so far as they correspond to an {\em independent} environment
acting on each qubit, then the answer to the question changes from
``nothing'' to ``quite a bit.''  For if the environments of each
qubit are independent, then irrespective of the form of those 
environments, the error operators $A_i$ of the complete system
become a direct product of error operators on individual qubits,
as in Eq. (\ref{bigarray}):
\begin{equation}
A_{\{i1,i2,i3,...\}}=A_{i1}^{(1)}\otimes A_{i2}^{(2)}\otimes 
A_{i3}^{(3)}\otimes ...\label{full}
\end{equation}
Now, each qubit error operator $A_{in}^{(n)}$ can be expanded in
terms of the matrices for the canonical error operators (since
they form a complete set):
\begin{equation}
A_{in}^{(n)}=\sum_{k=0}^3\alpha_{i,n,k}E_k,
\end{equation}
so that the full error operator can be expanded as a sum of products
of canonical error operators:
\begin{equation}
A_{\{i1,i2,i3,...\}}=\sum_{k1,k2,k3,...}\alpha_{\{kn\}}E_{k1}^{(1)}\otimes 
E_{k2}^{(2)}\otimes E_{k3}^{(3)}\otimes ...\label{bigex}
\end{equation}
Now, finally, it should be clear what good this is: if the error-correction
scheme is capable of correcting all the ``canonical'' errors which occur
in the expansion of the right-hand side of Eq. (\ref{bigex}), then, because
of the linearity of the error correcting process, the actual error operator
$A_{\{i1,i2,i3,...\}}$ will be successfully corrected.

This is still not the end of the story, because the terms in
Eq. (\ref{bigex}) will generally contain {\em all} possible products
of the $E$ operators, and we have argued that there cannot exist an
error correction procedure that corrects for all such errors.  In fact
we have to invoke one more physical requirement which we expect most
sensible environments to obey: at a function of the strength of
interaction between the system and the environment, or as a function
of the time of interaction, we expect that for weak coupling, or short
time, the dominant error operator should be very close to the
identity.  That is to say, the system cannot be corrupted very much
after a short period of interaction.  This principle allows us to
argue which terms will predominate in the sum Eq. (\ref{bigex}) for a
sufficiently weak coupling to the environment.  There will be one of
the $A$ operators in Eq. (\ref{bigex}) that is close to the identity
operator, which is $0^{th}$ order in time $t$.  Then there will be a
set of $A$'s for which the leading terms in the sum of
Eq. (\ref{bigex}) are ``single-error'' terms, involving only one
non-identity $E$ operator; for a wide class of noisy
environments\cite{Chuang}, these will be of order $O(t^{1/2})$.  The
next group of $A$'s will be ``double-error'' terms, which will be
$O(t^{2/2})$; the ``triple-error'' terms will be $O(t^{3/2})$; and so
forth.  Thus, the generic error correcting code which corrects some
number of ``canonical'' errors of Eq. (\ref{bigarray}) will take care
of the most important parts of the generic errors described by $A$ (at
least, at early times).  The final statement (which I will make
without proof) is that if one has an $e$ error correcting code
(correcting canonical errors), then after correction the density
matrix $\rho(t)$ will differ from the ideal state $\Psi(t)$ by terms
of order $t^{e+1}$:
\begin{equation}
\langle\Psi(t)|\rho(t)|\Psi(t)\rangle=1-ct^{e+1}+...
\end{equation}
The constant $c$ in front of the error term will be dependent on the
details of the noise as expressed in the $A$'s in Eq. (\ref{bigex}).

\subsection{Simple Example of Error Correction}

Having laid out the general principles of error correction, I want to
briefly review how it actually works for the simplest ``canonical''
example, the one of five qubits subjected only to the one-bit standard
errors of Eq. (\ref{bigarray}).  In our initial work\cite{us} we
embarked on a purely numerical search for two orthogonal vectors
$|v_0\rangle$ and $|v_1\rangle$ in the 32-dimensional Hilbert space of
the five spins which would satisfy all the conditions
Eqs. (\ref{cond1}) and (\ref{cond2}).  We indeed succeeded by using
this strategy, but in fact there is a much better strategy that was
developed very rapidly by a number of other authors, which I will
indicate briefly here.  The good approach is a group-theoretic
one\cite{Gottes,Calder}: suppose there is a set of operators $M_i\in
M$ each of which leave the code vectors invariant:
\begin{equation}
M_i|v_0\rangle=|v_0\rangle,~~~~M_i|v_1\rangle=|v_1\rangle.\label{stab}
\end{equation}
We note that if such a set of operators exists it must form a
group, and that it is a reasonable guess that this group is Abelian
(i.e., all the $M$'s commuting).  
The group structure is guaranteed because if $M_1$ and $M_2$
both satisfy Eq. (\ref{stab}), then obviously $M_1M_2$ does as well.
The fact that these operators should also commute is not guaranteed, but
is suggested by Eq. (\ref{stab}); 
an elementary fact which we learn in quantum mechanics is that
commuting operators have simultaneous eigenvectors, and Eq. ({\ref{stab})
asserts that these operators have at least two
simultaneous
eigenvectors, $|v_0\rangle$ and $|v_1\rangle$.  (Commutivity would 
be assured if we could assert that the operators had all 32 eigenvectors
in common, not just 2.  This will turn out to be the case\cite{Gottes}.)

Now, the next reasonable guess to be made is that if these group elements
$M$ exist, they should themselves be expressible as products of the
canonical $E$ operators of Eq. (\ref{canon}).  After all, the $E$ operators
are basically spin-1/2 angular momentum operators (i.e., the Pauli matrices),
and these operators always either commute or anticommute.  So, we have
a chance of building a set of commuting $M$s this way.

In fact, the possibility that such operators may also anticommute is
the final piece of this mathematical trickery.  For suppose that we
have found a set of $M$s which commute with each other, but do not
commute with all the canonical error operators $A^{canon.}$ of
Eq. (\ref{bigarray}).  This means that there will be one operator
$A_i^{canon.}$ (or more) which an element $M_\alpha$ anticommutes with:
\begin{equation}
M_\alpha A_i^{canon.}=-A_i^{canon.}M_\alpha.
\end{equation}
But watch what happens when I take the matrix elements of this equation
between code vectors $v_{0,1}$:
\begin{equation}
\langle v_\alpha|M_\alpha A_i^{canon.}|v_\beta\rangle=
-\langle v_\alpha|A_i^{canon.}M_\alpha |v_\beta\rangle
\end{equation}
\begin{equation}
\langle v_\alpha|A_i^{canon.}|v_\beta\rangle=
-\langle v_\alpha|A_i^{canon.}|v_\beta\rangle=0.
\end{equation}
So, we see that the anticommuting condition leads to the satisfaction
of the error correction conditions Eq. (\ref{cond1}) and (\ref{cond2}),
where the first condition is satisfied with the extra condition that
both matrix elements are equal to zero.  The argument just given is
obviously not true for the case that $A_i^{canon.}$ is the identity
operator (since this will commute with everything); but in this case
$A_i^{canon.}$ is in the group $M$, which also leads to Eqs. (\ref{cond1})
(\ref{cond2}) being satisfied.

I can now finish this off by giving the full statement of the result
of Gottesman\cite{Gottes} and of Calderbank {\em et al.}\cite{Calder}:
{\bf Vectors $|v_\alpha\rangle$ can be corrected when subjected to
errors $A_i$ if they are the eigenvectors of an Abelian group of
operators $M$ such that every operator $A_i^\dagger A_j$ either 1) is
itself a member of the group $M$, or 2) anticommutes with at least one
element of $M$.}

This result has provided a very useful means of searching for and
discovering a vast variety of error correcting schemes.  The five-bit
error correcting code can be very succinctly expressed using this
language: the code vectors are eigenstates of the sixteen-element
Abelian group which is generated by the four operators:
\begin{equation}
\begin{array}{r}
E_1^{(1)}E_1^{(2)}E_3^{(3)}E_3^{(5)}\\
E_1^{(2)}E_1^{(3)}E_3^{(4)}E_3^{(1)}\\
E_1^{(3)}E_1^{(4)}E_3^{(5)}E_3^{(2)}\\
E_1^{(4)}E_1^{(5)}E_3^{(1)}E_3^{(3)}
\end{array}\label{syn}
\end{equation}
One choice of the two vectors which are simultaneous eigenvectors
of these operators, which may be found by standard projection-operator
techniques from group theory, are:
\begin{eqnarray}
|v_0\rangle &=& |00000\rangle\label{sym0}\\
&+& |11000\rangle +|01100\rangle +|00110\rangle +|00011\rangle +|10001\rangle
\nonumber \\
&-& |10100\rangle -|01010\rangle -|00101\rangle -|10010\rangle -|01001\rangle
\nonumber \\
&-& |11110\rangle -|01111\rangle -|10111\rangle -|11011\rangle -|11101\rangle
\nonumber
\end{eqnarray}
and
\begin{eqnarray}
|c_1\rangle &=& |11111\rangle\label{sym1}\\
&+& |00111\rangle +|10011\rangle +|11001\rangle +|11100\rangle +|01110\rangle
\nonumber \\
&-& |01011\rangle -|10101\rangle -|11010\rangle -|01101\rangle -|10110\rangle
\nonumber \\
&-& |00001\rangle -|10000\rangle -|01000\rangle -|00100\rangle -|00010\rangle
\nonumber .
\end{eqnarray}
The entire error-correction process can also be described very succinctly
and physically in this language:  The prescription is to perform a 
{\em measurement} of the value of each of the four generators of the
$M$ group, Eq. (\ref{syn}).  (Although I have not mentioned it previously,
they are in fact Hermitian operators.)  Being the product of four spin-1/2
operators, the measurement can only have two outcomes, $\pm 1$.  If the
measurement outcomes are all $+1$, this indicates that the state is free
from error, and nothing need be done.  It turns out that each other
fifteen patterns of $\pm 1$ in this measurement indicates which of the
fifteen one-error operators of Eq. (\ref{bigarray}) that the state has
been subjected to.  Knowing which of these (unitary) operators the
state has been subjected to, simply performing the inverse unitary
operation restores the state to its correct value.  

Both these multi-spin measurements and the final unitary correction
operations are within the capability of a quantum computer as
specified in the first part of this paper.  Error correction indicates
a modified paradigm for quantum computation, however; instead of
measurements being performed only at the final, ``readout'' phase of
the computation, error correction indicates that measurements should
be performed at fixed intervals throughout the entire course of the
computation.

\subsection{Final Remarks on Error Correction, etc.}

As a summary remark on all this, I freely admit that the foregoing
barely scratches the surface of the current activity in the theory of
quantum quantum computation, even of the theory of error correction in
quantum computation.  Any such survey must be obsolete as soon as it
is written, with new families of error correcting codes being
discovered, their connection with classical error correcting theory
being further uncovered, and proposals being advanced for how to use
error correction in a full protocol for performing reliable quantum
computation with faulty elements.  The claim which is presently being
evaluated is that if the error rate (including that induced by
interaction with the environment, as well as inaccuracies in the
implementation of the quantum gates) is below a certain value, then
reliable quantum computation of indefinite scale and duration becomes
possible.  The bad news in this is that this threshold for reliable
quantum computation is presently quite low, in the neighborhood of
$10^{-5}$ per ``clock cycle'' (intervals between quantum gate
operations).  So, there's clearly a lot more territory for the
theory to explore at this point, and it goes without saying that
the experimental situation is still in its infancy.  Perhaps some
fresh-minded youngster reading this will have a good idea for how
to make good progress in bringing quantum computation closer to
reality.

I am grateful to the Army Research Office for the support of this work,
as well as the hospitality of the Program on Quantum Computers and
Quantum Coherence at the Institute for Theoretical Physics at UCSB
where some of this work was completed.

\pagebreak
\begin{figure}

\vspace{1in}

\centerline{\epsfbox{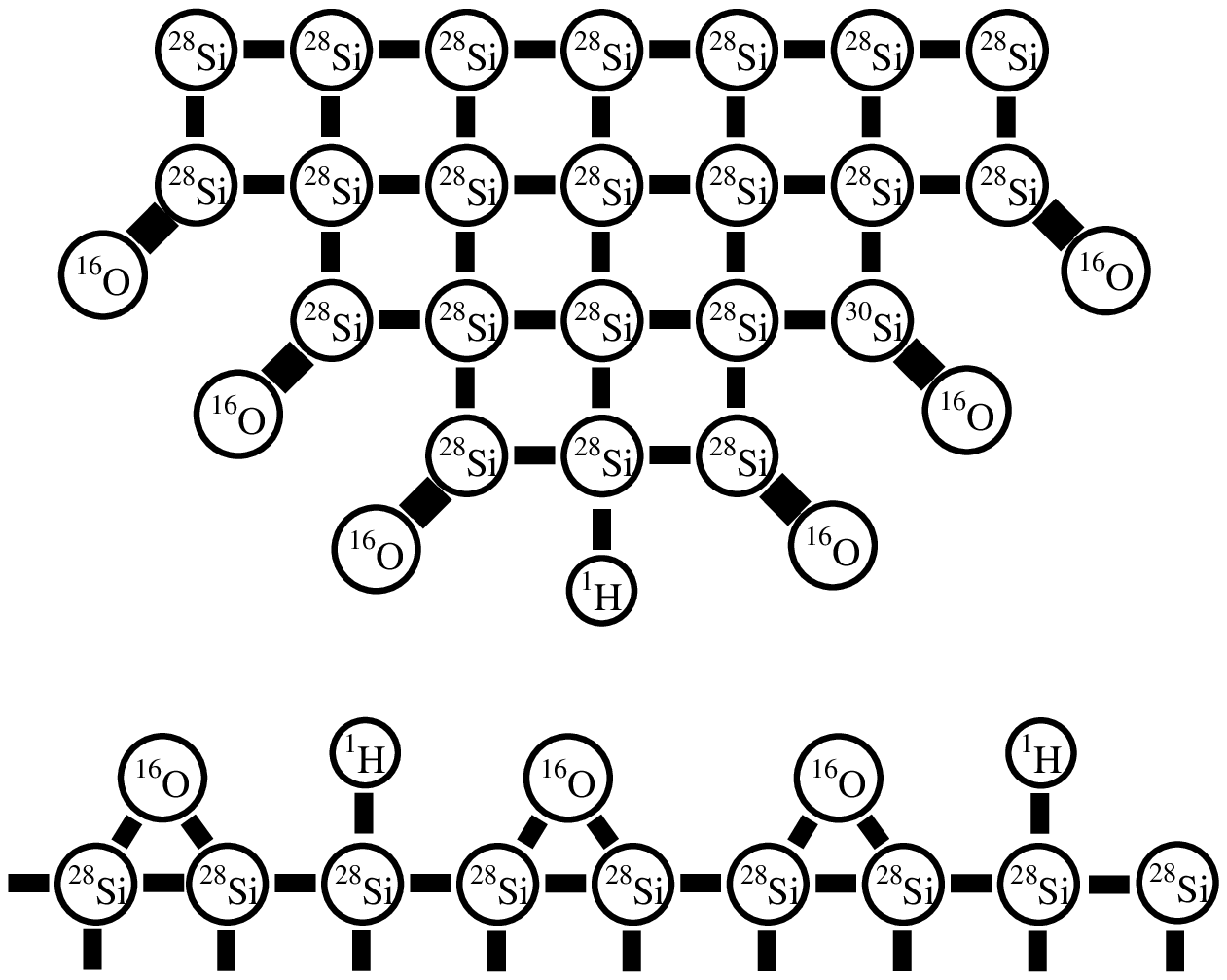}}

\vspace{1in}
\end{figure}
\end{document}